# COSMIC – The SLAC COntrol System MIgration Challenge

M. Clausen (DESY), Ron MacKenzie, Robert Sass, Hamid Shoaee, Greg White, Leeann Yasukawa
SLAC, Stanford, CA 94025, USA

Abstract

The current SLC control system was designed and constructed over 20 years ago. Many of the technologies on which it was based are obsolete and difficult to maintain. The VMS system that forms the core of the Control System is still robust but third party applications are almost non-existent and its long-term future is in doubt. The need for a Control System at SLAC that can support experiments for the foreseeable future is not in doubt. The present B-Factory or PEPII experiment is projected to run at least 10 years. An FEL laser of unprecedented intensity plus an ongoing series of fixed target experiments is also in our future. The Next Linear Collider or NLC may also be in our future although somewhat farther distant in time. The NLC has performance requirements an order of magnitude greater than anything we have built to date. In addition to large numbers of IOCs and process variables, Physicists would like to archive everything all the time. This makes the NLC Control System a bit like a detector system as well. The NLC Control System will also need the rich suite of accelerator applications that are available with the current SLC Control System plus many more that are now only a glimmer in the eyes of Accelerator Physicists. How can we migrate gradually away from the current SLC Control System towards a design that will scale to the NLC while keeping everything operating smoothly for the ongoing experiments?

## 1 ORACLE

Recent releases of the Oracle RDBMS provide expanded capabilities that make it reasonable to consider using it for configuration information, archive and perhaps even soft real-time data.

### 1.1 Configuration Information

Since its inception in the early 1980's, the SLC control system has been driven by a highly structured memory resident real-time database. While efficient, its rigid structure and file-based sources makes it difficult to maintain and extract relevant information. The goal of transforming the sources for this database into a relational form is to enable it to be part of a Control System Enterprise Database that is an integrated central repository for SLC accelerator device and control system data with links to other associated databases.

We have taken the concepts developed for the NLC Enterprise Database [1] and used them to create and load a relational model of the online SLC control system database. This database contains data and structure to allow querying and reporting on beamline devices, their associations and parameters. In the future this will be extended to allow generation of controls software configurations for EPICS and SLC devices, configuration and setup of applications and links to other databases such as accelerator maintenance, archive data, manufacturing history, cabling information, documentation etc. The database is implemented using Oracle 8i.

Figure 1: Proposed Configuration of SLC and EPICS

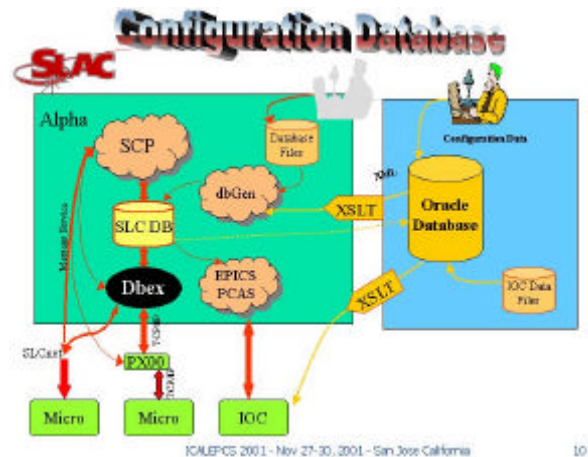

Real-time Databases Using Oracle

Figure 1 above depicts the use of Oracle for maintenance of configuration as we envision it to be. We presently have a substantial subset of SLC data in Oracle and are using it for some query applications but are not yet generating the SLC database or loading EPICS IOCs with data generated from Oracle.

### 1.2 Archive Data

Presently, the SLC and EPICS archive systems are completely separate. They each have disadvantages and neither will scale to the Petabyte sizes eventually needed by the NLC. Recent test results [2] for archiving into Oracle are very encouraging and present plans are to move forward rapidly on this part of the Cosmic development.

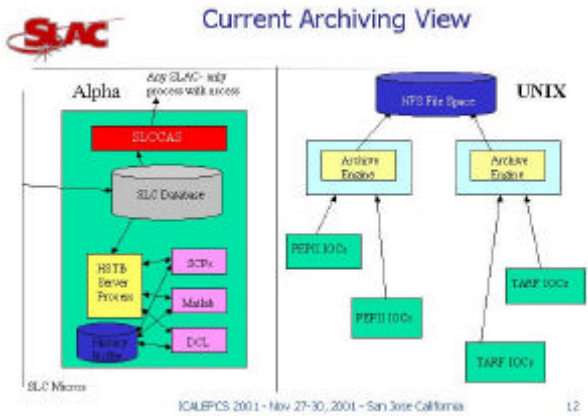

Figure 2: Current Archive Configuration in the SLC Control System

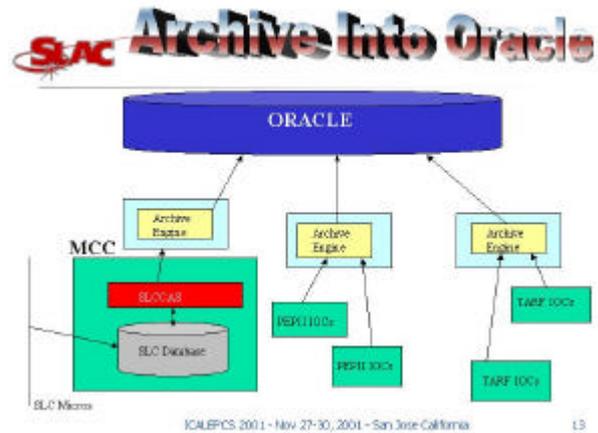

Figure 3: Proposed Archive Configuration for the SLC Control System

The SLC archive system has been running for many years and one can retrieve and analyze archive data over that entire span of time. Retrieval times are acceptable and data can be plotted using standard SLC Control System utilities. The data can also be exported to Matlab for detailed analysis and plotting. Despite these capabilities, it does have some serious disadvantages.

1. Data can only be retrieved and analyzed on the MCC VMS system.
2. There's a 6-minute update interval for most signals which is too coarse for many kinds of analysis.
3. Data is stored in an internal format and so must be explicitly exported by special software for use by other tools.

Figure 2 above shows an overview of the current SLC archive systems. You see how the two systems are completely separate without mutual access to data.

The EPICS archiver saves data using EPICS monitors and thus only when it changes or at some pre-defined maximum interval. It is supported by the EPICS collaboration and works with a number of different EPICS tools but it also has some serious disadvantages.

1. It uses a doubly linked set of files so all files must be online for it to work.
2. There are a limited number of tools for data access and maintenance.
3. Since the data are kept in binary files, it's difficult to use commercial tools to access/analyze the data.

Figure 3 above depicts the proposed solution to the archive problem. The Portable Channel Access Server running on the MCC VMS system will interface to an Archive Engine that will be used to funnel data into Oracle. Other EPICS IOCs in the system will also use other instances of the Archive Engine so all SLC Control System data will be in a common Oracle archive. Once in Oracle, a wide variety of Oracle and third-party tools can be used to retrieve and analyze the data.

## 2 AIDA

Higher-level applications need to access a logically related set of data that is in different data stores and may require different processing. Aida is envisioned to be a distributed service that allows applications access to this wide variety of Control System data in a consistent way that is language and machine independent. It has the additional goal of providing an object-oriented layer for constructing applications on top of multiple existing conventional systems like EPICS or the SLC Control System.

### 2.1 Overview

The client interface to Aida is via the CORBA Interface Definition Language or IDL. This interface is the "contract" between the clients and the Aida services and sources. Aida uses an Oracle database to map the names requested to services and sources. Initial discovery is made at runtime and subsequent requests can the go directly to the data, bypassing the Oracle access.

Requests can be either synchronous or asynchronous. The CORBA Event service is presently used for monitors. The CORBA Notify service is layered on top of the Event Service and has additional attractive features like quality of service client-side and event filters.

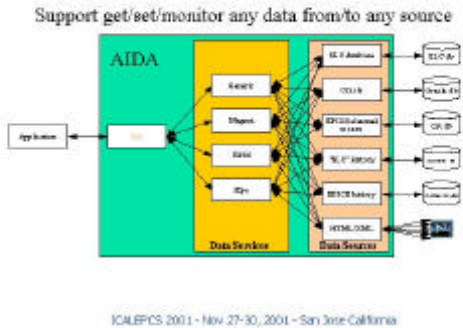

Figure 4: Aida Architecture Overview

Figure 4 above shows the major parts of the Aida architecture. The client's request is dispatched to the appropriate service which then gets the data form the appropriate source. The service may pre/post-process the data before storing/retrieving the requested data items. A generic service is provided for those data items that don't require additional processing.

*2.2 Request Details*

Figure 5 below follows a typical request through the Aida system.

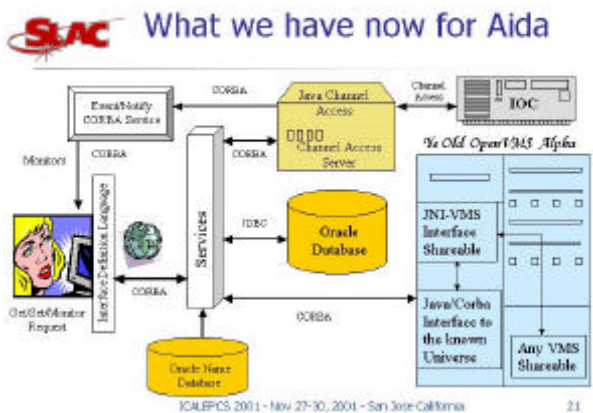

Figure 5: Aida Request Example

The client interfaces with an Aida object via the methods and structures specified in the IDL or Interface Definition Language. The method invocation on the remote CORBA object determines the service and sources of the data items requested by accessing the Oracle name database or its local cache if the item is still there from a previous request. The name database has information about the services and sources needed to process the request. Each of the data sources has one or more Aida servers to access the local data store. Requested data is either returned directly to the client in the case of a synchronous read or asynchronously using the Event service if the request is for a Monitor.

## 3 JOIMINT/MADAM

A final piece of the migration puzzle is a new display manager/user interface tool. The existing SLC interface is a now antiquated touch panel that interfaces with the underlying applications. It may be old but it's very fast and thus operators are able to use it for effective control of the accelerator. The EPICS DM (DM2K EDM) can only read/write to data channels but the SLC interface has several additional features that will be required in any new user interface.

- It communicates with underlying applications.
- It can capture keystrokes/operations and save them to a file. This file can be edited with basic loop and conditional controls and later replayed.
- Separate local and system message windows.
- Dynamic loading and configuration of new objects

Matthias Clausen of DESY started preliminary work on this and there is an update to be presented at this conference [3].

## 4 SUMMARY

Figure 6 below shows how using Oracle, Aida and JoiMint/Madam we can provide an environment for porting and implementing applications allowing us to gradually offload VMS while giving us a development environment that will hopefully be applicable to the NLC.

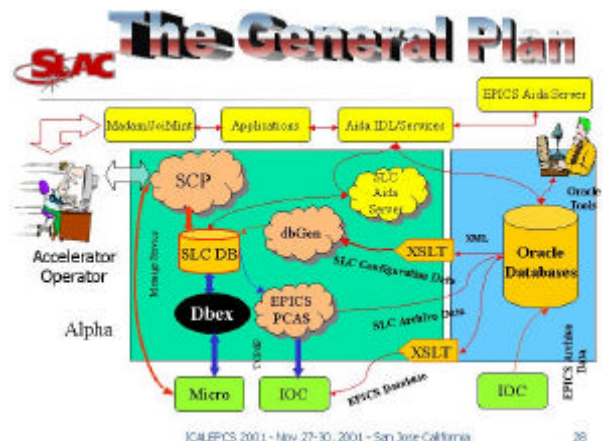

Figure 6 General Migration Plan

## 4.1 Archive to Oracle

Because early tests have been so promising, probably the first part of the migration will be the archive data. Merging both EPICS and SLC archive data in an Oracle database will give us good experience in managing large data stores. It will also provide an opportunity to explore the wide variety of Oracle and third-party tools for data extraction and analysis.

## 4.2 Configuration Management in Oracle

We have already done considerable work in modeling and implementing the SLC database in Oracle. Several other EPICS sites have used Oracle to store their EPICS database and we may use their implementation if it can be reasonably integrated with our existing design. The file system based storage of database information has served us well for many years so we'll need a high degree of confidence in the new system before switching over.

## 4.3 Aida

We plan to finish prototyping and start detailed design early next year. We hope to have a production quality version ready for initial application development later on in the year. CORBA capabilities are rapidly evolving. If you procrastinate long enough there will be a specification and eventually a product to do what you want. Current CORBA specifications already encompass real-time capabilities, failover and redundancy, messaging and embedded CORBA. As part of the telecommunications applications there is also a CORBA message logging service. After the core of Aida is robustly functional, we'll need to continuously monitor new CORBA capabilities and incorporate them or replace existing implementations as warranted.

## 4.4 JoiMint/Madam

This development is presently being undertaken by DESY based on requirements and initial prototyping work done at SLAC. An intelligent display capability is crucial to the Cosmic plan.

## 5 THE FUTURE

Even after the major applications have been ported and enhanced to the new environment, we have not provided for migrating the running database and the interfaces to our Intel Micros. Perhaps in the end these will just stay on a VMS system with the bulk of the applications running on the outside. These micros present several unique challenges.

- They run the iRMX operating system, which is now almost as obscure as VMS.
- Much of the code is tightly coupled to the Intel architecture in its handling of segments and pointers.
- All communication is done through the SLC home-built message system. We would either need to port support for it to other platforms or consider changing this to a CORBA message service.

None of the above options are particularly attractive and involve touching a lot of very old code that is at the heart of controlling the running of the accelerator. In the end, if VMS remains a supported operating system, we may retain the core of the database and communications on VMS while exporting the bulk of the applications and displays to a more distributed and modern environment.